\documentstyle[aps,prl,preprint]{revtex}

\newcommand{\ovl}{\overline}
\newcommand{\ul}{\underline}
\newcommand{\tr}{{\rm tr}\,}

\tightenlines

\title{Fluctuations and Ergodicity of the Form Factor of Quantum
Propagators and Random Unitary Matrices}
\author{Fritz Haake, Hans-J{\"u}rgen Sommers and Joachim Weber}
\address{Fachbereich Physik, Universit\"at-GH Essen\\
45117 Essen\\ Germany}

\begin{document}

\draft
\date{Date: \today}
\maketitle
\begin{abstract}
We consider the spectral form factor of random unitary matrices as well
as of Floquet matrices of kicked tops, as given by the (squared moduli
of) the traces $t_n={\rm Tr}F^n$ with the integer ``time'' $n=0,\pm
1,\pm 2,\ldots$. For a typical matrix $F$ the time dependence of the
form factor $|t_n|^2$ looks erratic; only after a local time average
over a suitably large time window $\Delta n$ does a systematic time
dependence become manifest. For matrices drawn from the
circular unitary ensemble we prove ergodicity: In the
limits of large matrix dimension and time window $\Delta n$ the local
time average has vanishingly small ensemble fluctuations and may be
identified with the ensemble average. By numerically diagonalizing
Floquet matrices of kicked tops with a globally chaotic classical limit
we find the same ergodicity. As a byproduct we find that the traces
$t_n$ of random matrices from the circular ensembles behave very much
like independent Gaussian random numbers. Again, Floquet matrices of
chaotic tops share that universal behavior. It becomes clear that the
form factor of chaotic dynamical systems can be fully faithful to
random-matrix theory, not only in its locally time-averaged systematic
time dependence but also in its fluctuations. \end{abstract}

\vspace{1 cm}


\section{Introduction}
In the present paper we propose to discuss the time dependence 
of the form factor of individual dynamical systems and of individual random
matrices and compare with the ensemble based predictions of random-matrix
theory. Our investigation was motivated by recent efforts to extract 
universality from semiclassical arguments. Among these, Smilansky and
coworkers \cite{SMI1,SMI2} have argued that random-matrix behavior should prevail in the 
autocorrelation function of the spectral determinant of dynamical
systems, at least after a suitable average; the conclusions were based on
assuming Gaussian statistics for the form factor fluctuations. Similarly
motivated work by Prange \cite{PRA} and Kunz \cite{KUN} has partial
overlap with ours.

In order to look for universal spectral
fluctuations from a given quasi-energy spectrum one must first 
``unfold'' the spectrum to constant mean level spacing, or else  
systematic system specific variations of the density of levels across
the spectrum would be mixed up with potentially universal fluctuations.
Moreover, smoothing over certain evergy intervals, denoted by the overbar
in $\ovl{\rho(E+\Delta E)\, \rho(E)}$, for the correlation
functions of the density of levels or over suitable time intervals for its
Fourier transform, the form factor, is necessary.
Without such averaging the form factor would
display erratic variations not at all resembling the smooth behavior predicted
by random-matrix theory for averages of matrix ensembles. 
The level density correlator would even consist of delta peaks.

If a conservative quantum system with global chaos in its classical limit
displays universal spectral fluctuations \`{a} la random-matrix theory,
such universality may reveal itself in the distribution $P(s)$ of nearest-neighbor
spacings,
for which random-matrix theory predicts the power law
$P(s) \sim s^\beta$ as $s \to 0$, and the degree $\beta$ of 
level repulsion is determined by symmetry, most importantly the 
absence or presence of time reversal invariance. Equally popular and even
more easily amenable to theoretical reasoning is the two-point function of the
level density $C(\Delta E) = \ovl{\rho(E+\Delta E)\, \rho(E)}- \ovl{\rho(E)}^2$, 
for which universal 
behavior implies the same power law for small energy difference
$\Delta E$.
The Fourier transform of that correlation function, the so-called form factor
$K(\tau)$, may of course also be used as an indicator of universality (or of
lack thereof).
Ergodicity of the underlying classical dynamics \cite{HOdA} imparts a
linear dependence on the time $\tau$, $K(\tau) \sim |\tau|$, for
$|\tau|$ larger than the shortest periods of classical periodic orbits,
but smaller than the Heisenberg time $\tau_H$ (the time scale
corresponding to the mean level spacing as a unit of energy). For larger
times, $|\tau| \gg \tau_H$, the form factor tends to a constant value,
and the character of the approach is related to the degree of level
repulsion.

For the sake of concreteness we base our work on unitary $N \times N$ matrices,
such as arise as Floquet matrices $F$ for periodically driven systems with a
compact phase space and thus finite dimensional Hilbert space.
As examples  of such dynamics we shall take kicked tops, ones with the 
spherical phase space dominated by chaos and, on the other hand, symmetries
chosen so as to fit any of the three principal universality classes 
(orthogonal, unitary and symplectic).
The corresponding random-matrix ensembles are, of course, the so-called 
circular ones of Wigner and Dyson (COE, CUE, CSE).

Our findings within random-matrix theory extend the ones previously presented
in \cite{HKS} and may be summarized as follows. They all concern the ``traces''
$t_n = \tr F^n$ with $n = 1, 2, \dots$, which give the form factor as
$K(n) = |t_n|^2$; the integer exponent $n$ serves as a discrete time,
counting the number of periods of the external driving. (In the
large-$N$ limit we shall eventually introduce $\tau=n/N$ as a
quasi-continuous time.)
The circular ensembles yield marginal probability densities
$\langle \delta(t-t_n) \rangle$ which, in the limit of large matrix dimensions,
$N\to \infty$, assign Gaussian statistics to finite-order moments. In particular,
the first four moments bear relations of Gaussian character, as if coming from
$P(t) = (\pi \ovl{ |t_n|^2} )^{-1} \exp \left( -|t|^2 / 
\ovl{ |t_n|^2 } \right) $.
For different traces $t_n$ and $t_m$ we show, again for the large-$N$ limit,
that the unitary circular ensemble does not give cross-correlations, in the sense
$\ovl{|t_n|^2 |t_m|^2} /(\ovl{|t_n|^2}\; \ovl{|t_m|^2}) - 1 
= {\cal O}(1/N)$ for $m \ne n$.  
Using these results for ensemble averages we show for the CUE that the
form factor is ergodic in the large-$N$ limit: The  time average
$\langle |t_n|^2/\ovl{|t_n|^2} \rangle = (\Delta n)^{-1} \sum_{n'=n}^{n+\Delta n} 
|t_{n'}|^2 /\ovl{|t_n'|^2}$ has an ensemble variance vanishing in the limit of a 
large temporal window $\Delta n$ as $1/ \Delta n$.
This means that with overwhelming probability every random unitary matrix drawn from
the appropriate circular ensemble has the same time averaged form factor
and that the latter equals the ensemble-averaged form factor. Inasmuch
as a dynamical system has a Floquet matrix typical for the appropriate ensemble one can
expect universality for its time averaged form factor as well.

By numerically diagonalizing Floquet matrices of kicked tops from the various 
universality classes we have calculated, for each of these, the form factor
$K(n) = |t_n|^2$ and its time average over a finite window $\Delta n$. 
Normalizing to the ensemble-averaged form factor $\ovl{|t_n|^2}$ \`{a} la
random-matrix theory we throw all $|t_n|^2/ \ovl{|t_n|^2} $ for a given top into
a histogram and find this to reproduce the Gaussian behavior predicted by 
random-matrix theory. The ergodic character of random matrices is also 
respected in full by the Floquet matrices of chaotic tops: The time-averages
$\langle |t_n|^2/\ovl{|t_n|^2} \rangle$ come out to have variances within the 
respective data sets $\{ t_n \}$ varying with the time window $\Delta n$ as
$1/\Delta n$. 

\section{The Form Factor and its Fluctuations in Ensembles of Random  Matrices}

The density of eigenvalues ${\rm e}^{-{\rm i}\varphi_i}$ of a unitary matrix
$F$ can be written as
\begin{equation}
\rho(\varphi) = \frac{1}{2 \pi N} \sum_{i=1}^N 
\sum_{n=-\infty}^\infty {\rm e}^{{\rm i}n (\varphi-\varphi_i)} =
\frac{1}{2 \pi N} \sum_{n=-\infty}^\infty t_n {\rm e}^{{\rm i} n \varphi} \; .
\label{eq1}
\end{equation} 
We herein meet the traces
\begin{equation}
t_n = \sum_{i=1}^N {\rm e}^{-{\rm i} n \varphi_i} = \tr F^n, \quad n = 0, \pm1,
\pm2, \dots \; .
\label{eq2}
\end{equation} 
If $F$ is drawn from a homogeneous ensemble of random-matrices like any of the 
familiar circular ones of Wigner and Dyson, one has the ensemble average 
$\ovl{ t_n t_{n'} } = \delta_{n,-n'} \ovl{|t_n|^2}$ and thus the two-point correlation 
function
\begin{eqnarray}
C(e) & = & \ovl{\rho(\varphi + e 2 \pi/N) \rho(\varphi)} - \ovl{\rho}^2  \label{eq3}\\
     & = & \sum_{n \ne 0 } \ovl{|t_n|^2} {\rm e}^{{\rm i} 2 \pi n e/N} \, . \nonumber
\end{eqnarray}
The mean form factor $|t_n|^2$ is well known for the circular ensembles \cite{MET,HKS}
and will be given further below.
Since we want to determine higher-order moments like $\ovl{ |t_n|^4}$, it is best
to consider the characteristic function
\begin{eqnarray}
\tilde{P}_{n N}^\beta(k) & = & \overline{\exp\left(-\frac{{\rm i}}{2}
\sum_i (k {\rm e}^{-{\rm i}n\varphi_i} + k^* {\rm e}^{{\rm i} n \varphi_i})\right)}
\label{eq4} \\
& = & \overline{\prod_{i=1}^N \exp(-{\rm i}|k|\cos(n\varphi_i))} \; , \nonumber
\end{eqnarray}
where the joint densities of eigenphases of the circular ensembles \cite{MET,HKS}
may be used to calculate the averages for $\beta = 1$ (Circular Orthogonal Ensemble),
$\beta = 2$ (Circular Unitary Ensemble) and $\beta =4$ (Circular Symplectic Ensemble).
The homogeneity of these ensembles entails the characteristic function to depend on
$k$ and $k^*$ only through the modulus $|k|$ which we shall simply denote by
$k$ henceforth. It follows that all odd-order moments vanish, as well as those 
even-order ones where a trace $t_n$ is not accompanied by its complex conjugate 
$t_n^*$ to form powers of $|t_n|^2$. The non-vanishing moments are 
\begin{equation}
\overline{|t_n|^{2m}} = (-4)^m {2m \choose m}^{-1} 
\frac{\mbox{d}^{2m}}{\mbox{d}k^{2m}} \tilde{P}_{n N}^\beta(k) |_{k=0} \; .
\label{eq5}
\end{equation}

\section{Circular Unitary Ensemble}
The characteristic function is easily evaluated using any of the standard methods
of random-matrix theory and takes the form of a Toeplitz determinant,
\begin{equation}
\tilde{P}_{n N}^2(k)  =  \det M\, , \quad M_{\mu\nu}  =
 \sum_{s=-\infty}^{\infty} 
(-{\rm i})^{|s|} \, J_{|s|}(k) \, \delta_{\mu,\nu-ns} \; ,
\label{eq6}
\end{equation}
with $J_s$ the Bessel function of integer order. The desired Taylor coefficients of
$\det M(k)$ are most conveniently evaluated through the expansions
\begin{eqnarray}
M(k) & = & 1 + k M^{(1)} + \frac{k^2}{2} M^{(2)} + \frac{k^3}{3!} M^{(3)} + \dots \; ,
\label{eq7} \\
\det M(k) & = & {\rm e}^{\tr \ln M(k)} = \exp (\frac{k^2}{2} \tau_2 + \frac{k^4}{6!} \tau_4 +
\dots) \;\; . \nonumber
\end{eqnarray}
The derivatives $M^{(i)} = {\rm d}^i M(k)/{\rm d}k^i|_{k=0}$ can be calculated 
with the help of $J_s(0) = \delta_{s,0}$ and 
$\mbox{d} J_s(k)/\mbox{d}k = (J_{s-1}(k)-J_{s+1}(k))/2$ as
\begin{eqnarray}
M^{(1)}_{\mu \nu} & = & -\frac{{\rm i}}{2} (\delta_\mu^{\nu-n} + \delta_\mu^{\nu+n}) \;,
\label{eq8} \\
M^{(2)}_{\mu \nu} & = & -\frac{1}{4} (\delta_\mu^{\nu-2n} + \delta_\mu^{\nu+2n}) 
- \frac{1}{2} \delta_\mu^\nu \; , \nonumber \\
M^{(3)}_{\mu \nu} & = & \frac{{\rm i}}{8} (\delta_\mu^{\nu-3n} + \delta_\mu^{\nu+3n}) +
\frac{3 i}{8} (\delta_\mu^{\nu-n} + \delta_\mu^{\nu+n}) \; , \nonumber \\
M^{(4)}_{\mu \nu} & = & \frac{1}{16} (\delta_\mu^{\nu-4n} + \delta_\mu^{\nu+4n})
+ 
\frac{1}{4} (\delta_\mu^{\nu-2n} + \delta_\mu^{\nu+2n}) + \frac{3}{8}
\delta_\mu^\nu \; . \nonumber
\end{eqnarray}
Inasmuch as the Taylor coefficients of $\det M(k)$ are related to the moments
$\ovl{|t_n|^{2m}}$ and the Taylor coefficients of its logarithm 
$\ln \det M = \tr \ln M$ to the corresponding cumulants, their interrelations
\begin{eqnarray}
\tau_1 & = & \tr M^{(1)} \; , \label{eq9} \\
\tau_2 & = & \tr (M^{(2)} - M^{(1) \, 2}) \; , \nonumber \\
\tau_3 & = & \tr (M^{(3)} + 2 M^{(1)\,3} - 3 M^{(1)}M^{(2)}) \; , \nonumber \\
\tau_4 & = & \tr (M^{(4)} + 12 M^{(1)\,2} M^{(2)}
-6 M^{(1)\,4} - 3 M^{(2)\,2} -4 M^{(1)} M^{(3)}) \;  \nonumber
\end{eqnarray}
resemble the familiar ones between moments and cumulants of random variables.
One arrives at $\tau_1 = \tau_3 = 0$ and
\begin{eqnarray}
\tau_2 & = & \left\{
 \begin{array}{l@{\quad\quad}l}
  -n/2 & 0<n \le N \\
  -N/2 & N \le n \quad\quad,
 \end{array}
 \right. \label{eq11} \\
\tau_4 & = & \left\{
 \begin{array}{l@{\quad\quad}l}
  0 & 0<n \le N/2 \\
  \frac{3}{8} N-\frac{3}{4} n  & N/2 \le n \le N \\
  -\frac{3}{8} N & N \le n \quad\quad\quad\; ,
 \end{array}
 \right.
\label{eq12}
\end{eqnarray}
and these give the well-known CUE form factor
\begin{equation}
\ovl{|t_n|^2} = \left\{
 \begin{array}{l@{\quad\quad}l}
  n & 0<n \le N \\
  N & N \le n 
 \end{array}
 \right.
\label{eq13}
\end{equation}
as well as its variance
\begin{eqnarray}
 \mbox{Var}(|t_n|^2) & = & \ovl{|t_n|^4} - \ovl{|t_n|^2}^2 = 4
\tau_2^2+\frac{8}{3}\tau_4  \label{eq14} \\ 
  & = & \left\{
 \begin{array}{l@{\quad\quad}l}
  n^2 & 0<n \le N/2 \\
  n^2-2n+N & N/2 \le n \le N \\
  N^2-N & N \le n
 \end{array}
 \right. \\
 & = & \ovl{|t_n|^2}^2 + \ovl{|t_{2n}|^2} - 2\ovl{|t_n|^2} \; .
\label{eq15}
\end{eqnarray}
Within the interval $0<n \le N/2$ the mean $\ovl{|t_n|^2}$ and the variance
$\mbox{Var}(|t_n|^2)$ are related as if $t_n$ had a Gaussian distribution
\begin{equation}
P(t_n) = \frac{1}{\pi \ovl{|t_n|^2}} e^{-|t_n|^2/\ovl{|t_n|^2}} \; ,
\label{eq16}
\end{equation}
and that distribution also correctly reflects the vanishing of all odd-order
moments and cumulants. In \cite{HKS} we had established the more general
result that (\ref{eq16}) even reproduces all moments $\ovl{|t_n|^{2 m}}$
with $n \le N/m$. Below we shall show that the Gaussian relation between
$\ovl{|t_n|^2}$ and $\mbox{Var}(|t_n|^2)$ prevails even for $n > N/2$, up to
asymptotically negligible corrections of order $\ln(N)/N$.

\section{Circular Orthogonal and Symplectic Ensemble}
For the COE and CSE the evaluation of $P_{n N}^\beta (k) $ in (\ref{eq4})
proceeds analogously \cite{QSC2} and yields the Pfaffians
$P_{n N}^\beta (k) = \sqrt{\det A^\beta (k)}$,
\begin{eqnarray}
A^1_{\ul{m} \ul{m}'}(k) & = & \sum_{s,s'} J_{|s|}(k)\, J_{|s'|}(k)\,
\frac{(-i)^{|s|+|s'|}}{\ul{m}+ns} \, \delta_{\ul{m}}^{-\ul{m}-n(s+s')} \; , \label{eq17}\\
A^4_{\ul{m} \ul{m}'}(k) & = & (\ul{m}-\ul{m}') \sum_s J_{|s|}(2k)\,(-i)^{|s|}\,
\delta_{\ul{m}}^{-\ul{m}-ns} \; , \nonumber
\end{eqnarray}
wherein the underlined indices $\ul{m},\ul{m}'$ are semi-integer and in the ranges
$|\ul{m}|,|\ul{m}'| \le (N-1)/2$ for $\beta=1$ and $|\ul{m}|,|\ul{m}'| \le (2N-1)/2$
for $\beta = 4$. 
One encounters more complicated expressions for the expansion of
$\ln A^\beta(k)$ since $A^\beta(0) \ne 1$ and an additional factor $1/2$ in the 
$\tau_i$ due to the square root in the Pfaffian. 
To present our results in a concise form we introduce the shorthands
\begin{equation}
f_a^b = \sum_{m=a+1}^{b} \frac{1}{m+n-(N+1)/2} \, , \;\;
g_a^b = \sum_{m=a+1}^{b} \frac{1}{(m+n-(N+1)/2)^2} \; .
\label{eq18}
\end{equation}
For the COE we find the form factor
\begin{equation}
\ovl{|t^{\rm \small COE}_n|^2} = \left\{
 \begin{array}{l@{\quad\quad}l}
  2n - n f_{N-n}^N & 0<n \le N \\
  2N - n f_0^N & N \le n \quad\quad ,
 \end{array}
 \right.
\label{eq19}
\end{equation}
and the variance
\begin{eqnarray}
\mbox{Var}(|t^{\rm \small COE}_n|^2) -  \ovl{|t^{\rm \small COE}_n|^2}^2 \quad
\quad & = &  \label{eq20} \\ 
 & &  \mkern-220mu \left\{
\begin{array}{l@{\quad}l}
 8n(f_{-n}^N-\frac{1}{2}f^N_{N-n}-\frac{1}{4}f^{N+n}_{N-n})-2 n^2 g^N_{N-n} 
 & 0<n \le N/2 \\
 8n(f_{-n}^N+f_{-n}^n-\frac{1}{2}f_{N-n}^N-\frac{1}{4}f_{N-n}^{N+n})
 -2n^2g_{N-n}^N + 8 N - 2n & N/2 \le n \le N \\
 8n(f_0^N+\frac{1}{4}f_0^n-\frac{1}{4}f_N^{N+n})-2n^2 g_0^N-8N &
N\le n \;\;\; .
 \end{array}
\right.
\nonumber
\end{eqnarray}
For the CSE, i.e. the case $\beta=4$, Kramers' degeneracy is present. It suffices
to take every eigenvalue into account once: Assuming the matrix to be of size $2N$
we work with the ``trace'' $t_n^{\rm \small CSE} = \sum_{i=1}^{N}
{\rm e}^{-{\rm i} \varphi_i}
= \frac{1}{2} \tr F^n$ and thus account for every Kramers' doublet once.

The form factor and its variance are obtained as
\begin{equation}
\ovl{|t^{\rm \small CSE}_n|^2} = \left\{
 \begin{array}{l@{\quad\quad}l}
  n/2 - \frac{n}{4}  f_{-n-N/2}^{-N/2}  & 0<n \le 2N \\
  N & 2N \le n \quad\quad ,
 \end{array}
 \right.
\label{eq21}
\end{equation}

\begin{eqnarray}
\mbox{Var}(|t^{\rm \small CSE}_n|^2) - \ovl{|t^{\rm \small CSE}_n|^2}^2 \quad  & = & \label{eq22} \\
 & &   \mkern-200mu \left\{
 \begin{array}{l@{\quad\quad}l}
 -\frac{n^2}{8}g_{-n-N/2}^{-N/2}+\frac{n}{8}(f_{-n-N/2}^{-N/2}-f_{-N/2}^{n-N/2}) & 0<n\le N\\
 -\frac{n^2}{8}g_{-N/2}^{-n+3/2N} +\frac{n}{4}f_{-n-N/2}^{-N/2}+N-n & N\le n \le 2N\\ 
-N & 2N \le n \;\;\; .
 \end{array}
 \right. \nonumber 
\end{eqnarray}
As the important result we find that fluctuations are of the same order as the form factor
itself. We therefore cannot expect an individual Floquet or random
matrix to yield a sequence $|t_n|^2$ in accordance with the
ensemble mean without any averaging. We shall presently
show that the form factor is ergodic, i.e. the ensemble mean is 
approached by taking a local time average in Pandey's sense
\cite{PAN}.
Before proceeding to that endeavor we consider the fluctuations in the limit $N\to \infty$,
where they simplify significantly. 

\section{Asymptotic Fluctuations}
Prior to taking fluctuations to the limit $N\to \infty$ a 
normalization is in order. It seems quite natural to refer the discrete time $n$
to $N$, which is the Heisenberg time for $\beta = 1,2$ and half of
the Heisenberg time for $\beta = 4$. We write $\tau = n/N$, imagine $\tau$ kept
fixed as $N \to \infty$, and eventually allow $\tau$ to range among the real numbers.
The asymptotic form factors arising in that limit are of course well known. 

No simplification arises in the unitary case,
\begin{equation}
\frac{\ovl{|t_n|^2}}{N} \; \to \; K_{\rm \small CUE}(\tau) =  \left\{
 \begin{array}{l@{\quad\quad}l}
|\tau| & 0<|\tau| \le 1\\
 1 & 1\le |\tau| \end{array}
 \right.\; .
\label{eq23}
\end{equation}
The CUE form factor is everywhere continuous but its first derivative jumps at
$\tau = 0$ and at the Heisenberg time, $\tau = \pm 1$.

In the orthogonal and symplectic cases the sums $f_a^b$ and $g_a^b$ must be turned
into integrals and obviously yield logarithms. The COE form factor then becomes
\begin{equation}
\frac{\ovl{|t_n|^2}}{N} \; \to \; K_{\rm \small COE}(\tau) =  \left\{
\begin{array}{l@{\quad\quad}l}
 2 |\tau| - |\tau| \ln(2 |\tau|+1) & 0<|\tau| \le 1\\
 2- |\tau| \ln(\frac{2 |\tau| +1}{2 |\tau| -1}) & 1\le |\tau| \end{array}
 \right.\; .
\label{eq24}
\end{equation}
We still encounter a jump of $K'_{\rm \small COE}(\tau)$ at $\tau = 0$ but at the
Heisenberg time $\tau = \pm 1$ a jump arises only for the third derivative.

The asymptotic CSE form factor,
\begin{equation}
\frac{\ovl{|t_n|^2}}{N} \; \to \; K_{\rm \small CSE}(\tau) =  \left\{
\begin{array}{l@{\quad\quad}l}
 \frac{|\tau|}{2} - \frac{|\tau|}{4} \ln(|1-|\tau||) & 0<|\tau| \le 2\\
 1& 2\le |\tau| \end{array}
 \right. \;,
\label{eq25}
\end{equation}
displays a logarithmic singularity at $\tau = \pm 1$, i.e. at half the Heisenberg time.
Jumps arise in $K'_{\rm \small CSE}(\tau)$ at $\tau=0$ and in $K'''_{\rm \small CSE}(\tau)$
at the Heisenberg time $\tau = \pm 2$.

As for the variances ${\rm Var}(|t_n|^2/N) $ we find strictly Gaussian behavior in the limit 
$N \to \infty$,
\begin{equation}
\lim_{N \to \infty} \mbox{Var}(\frac{|t_n|^2}{N}) = \lim_{N \to \infty} \left(
\frac{\ovl{|t_n|^2}}{N} \right)^2 = K(\tau)^2 \; ,
\label{eq26}
\end{equation}
for all three circular ensembles. A little side remark is indicated, however, for the case of the
CSE. We here find a funny correction to the Gaussian behavior which according to taste one 
might keep or drop,
\begin{equation}
\lim_{N \to \infty} \mbox{Var}(\frac{|t_n|^2}{N}) - K_{\rm \small CSE}(\tau)^2 =
-\delta_{|\tau|, 1} \frac{\pi^2}{16} \; ,
\label{eq27}
\end{equation}
where $\delta_{|\tau|, 1}$ denotes the Kronecker delta. Inasmuch as the r.h.s. for $|\tau|=1$
(at half the Heisenberg time) is of order unity one would want to keep the correction and
even feast on its non-Gaussian character; on the other hand, if one wants to look at $\tau$
as a continuous variable the exceptional points $\tau = \pm 1$ accomodate removable singularities
without weight for integrals. Furthermore, the r.h.s. of (\ref{eq27})
is relatively small since $K_{\rm \small CSE}(\tau)$ diverges
logarithmically at $\tau = 1$.
Figure (\ref{fig1}) gives an impression for the approach of 
$\mbox{Var}(|t_n|^2/N) - (\ovl{|t_n|^2}/N)^2$ to the asymptotic $- \delta_{|\tau|,1} \pi^2/16$ as
$N$ grows. If one wants to study that approach analytically, one must isolate in  
$\mbox{Var}(|t_n|^2/N) - (\ovl{|t_n|^2}/N)^2$ the term which does not manifestly vanish for 
$N \to \infty$ at least as $\ln(N)/N$.
With the help of (\ref{eq21}) and (\ref{eq22}) one finds that term as 
\begin{equation}
h(n,N) = \left\{
 \begin{array}{l@{\quad\quad}l}
 -\frac{n^2}{8 N^2} g_0^N& 0<\tau\le 1\\
 -\frac{n^2}{8 N^2} g_n^{2N} & 1\le \tau \le 2 \quad ,
\end{array}
 \right. 
\label{eq28}
\end{equation}
which can be expressed in terms of the polygamma function $\psi^{(1)}$,
the first derivative of the digamma function, as 
\begin{equation}
h(n,N) = \left\{
 \begin{array}{l@{\quad\quad}l}
 -\frac{n^2}{8 N^2} (\psi^{(1)}(N-n+\frac{1}{2})- \psi^{(1)}(N+\frac{1}{2})) &
0<n\le N\\
 -\frac{n^2}{8 N^2} (\psi^{(1)}(n-N+\frac{1}{2})- \psi^{(1)}(N+\frac{1}{2}))  &
N\le n \le 2N \quad . 
 \end{array}
 \right. 
\label{eq29}
\end{equation}
The polygamma function $\psi^{(1)}$ has the familiar integral representation
\begin{equation}
\psi^{(1)}(z) = \int_0^\infty {\rm d} t \frac{t}{1-{\rm e}^{-t}}
{\rm e}^{-zt} \, .
\label{eq30}
\end{equation}
We need the foregoing representation for $z= N(1+1/(2N)), \, z = N(1-\tau+1/(2N))$
and $ z = N(\tau-1+1/(2N))$.
Obviously now, for $\tau \ne 1$ fixed and $N \to \infty$ we have
$\psi^{(1)} \to 0$ in all terms involved; at $\tau = 1$, however, the limit
$N \to \infty$ yields $\psi^{(1)}(1/2) = \pi^2/2$ and thus 
$h(N,N) = -\pi^2/16$. The decay of $h(\tau N,N)$ to zero for $N \to \infty$
at fixed $\tau \ne 1$ is found from the asymptotic large-$z$ behavior of the
polygamma function
\begin{equation}
\psi^{(1)}(z) = \frac{1}{z} + {\cal{O}} (\frac{1}{z^2})
\label{eq31}
\end{equation}
as
\begin{equation}
h(\tau N,N) = \left\{
 \begin{array}{l@{\quad\quad}l}
 -\frac{1}{8 N} \frac{\tau^3}{1-\tau} & 0<\tau< 1\\
 -\frac{1}{8 N} \frac{\tau^3}{\tau-1}  & 1< \tau \le 2 \quad .
 \end{array}
 \right.
\label{eq32}
\end{equation}

\section{Ergodicity of the Form Factor}
Fluctuations of the form factor tend to be suppressed by a local time average,
\begin{equation}
\langle |t_n|^2 \rangle = \frac{1}{\Delta n} \sum_{n'=n-\Delta n/2}^{n +\Delta n/2} |t_{n'}|^2 \; .
\label{eq33}
\end{equation}
As one increases the time window $\Delta n$ one expects the time average to become
equivalent to the ensemble average. Inasmuch as the systematic time dependence should not be
washed out by the time average, one must first let $N \to \infty$ and subsequently $\Delta n \to \infty$.

To uncover the expected ergodicity we consider the form factor as normalized to its ensemble
mean, $|t_n|^2/\ovl{|t_n|^2}$. We realize that 
$\ovl{\langle |t_n|^2/\ovl{|t_n|^2}\rangle} = 1$ holds trivially 
and propose to show that the ensemble variance of the temporal mean vanishes,
$ \mbox{Var}(\langle |t_n|^2/\ovl{|t_n|^2} \rangle ) \to 0 $, as $\Delta n \to \infty$.
The variance in question is defined as 
\begin{equation}
\mbox{Var} \left( \langle |t_n|^2/\ovl{|t_n|^2}\rangle \right) = 
\left( \frac{1}{\Delta n} \right)^2 \sum_{n',n'' = n-\Delta n/2}^{n+\Delta n/2}
\left( \frac{\ovl{|t_{n'}|^2 |t_{n''}|^2}}{\ovl{|t_{n'}|^2} \; \ovl{|t_{n''}|^2}}
-1\right) \; .
\label{eq34}
\end{equation}
Its evaluation requires knowledge of the cross-correlator $\ovl{|t_{n'}|^2 |t_{n''}|^2}$ for $n' \ne n''$.
We may determine the cross-correlator from the characteristic function
\begin{equation}
\tilde{P}(k_{n'},k_{n''}) = \ovl{\exp \left( - \frac{{\rm i}}{2} \sum_{i}
\left( ( k_{n'} {\rm e}^{-{\rm i} n' \varphi_i} + k_{n''} {\rm
e}^{-{\rm i} n'' \varphi_i}) + \mbox{c.c.}
\right) \right) }
\label{eq35}
\end{equation}
which generalizes the single-trace one (\ref{eq4}) and yields
\begin{equation}
\ovl{|t_{n'}|^2 |t_{n''}|^2} = (2 {\rm i})^4 \left. \frac{\partial^4}{\partial k_{n'} \partial k_{n'}^*
\partial k_{n''} \partial k_{n''}^* } \tilde{P}(k_{n'}, k_{n''})  \right|_0 \; .
\label{eq36}
\end{equation}
Once more employing the technique used to establish the single-trace characteristic function
one finds for the CUE
\begin{eqnarray}
\tilde{P}^2(k_{n'},k_{n''}) & = & \det M \; , \label{eq37} \\
 M_{\mu\nu} &  = &
\int_0^{2\pi}\frac{\mbox{d}\phi}{2\pi} {\rm e}^{{\rm i}(\mu-\nu)\phi}
\exp\left( -\frac{{\rm i}}{2}
(k_{n'} {\rm e}^{-{\rm i}n'\phi} + k_{n''} {\rm e}^{-{\rm i}{n''}\phi}+ \mbox{c.c.}) \right) \; ,
\nonumber
\end{eqnarray}
and from here $(n'>n'')$
\begin{equation}
\frac{\ovl{|t_{n'}|^2 |t_{n''}|^2}}{\ovl{|t_{n'}|^2} \; \ovl{|t_{n''}|^2}} -1  =
\left\{
 \begin{array}{l@{\quad\quad}l}
 \frac{N-n'-n''}{n' n''} & n' < N,\, n'+n'' > N \\
 -\frac{N-n'+n''}{Nn''} & N<n',\, n'-n''<N \\
 0 & \mbox{otherwise}\; . \end{array}
 \right. 
\label{eq38}
\end{equation}
Upon inserting these cross-correlations into (\ref{eq34}) the desired variance
is obtained.
An upper bound for this variance is found by neglecting all negative 
contributions to the sum, i.e. all terms with $n'\ne n''$. The estimate then
reads
\begin{equation}
\mbox{Var}\left( \langle |t_n|^2/\ovl{|t_n|^2}\rangle \right) \le 
\frac{1}{\Delta n} \stackrel{\Delta n \to \infty}{\longrightarrow} 0 \; .
\label{eq39}
\end{equation} 
The asserted ergodicity of the form factor is thus proven.

\section{Asymptotic Independence of the Traces}
With the cross-correlations (\ref{eq38}) at hand we can now show
that the traces $t_{n'}, t_{n''}$ with $n' \ne n''$ are asymptotically 
uncorrelated in the limit $N \to \infty$.
Again we assume $n'>n''$ and  
first consider the case $n'<N, \, n'+n''>N$ in (\ref{eq38}),
where we can estimate
\begin{equation}
\left| \frac{N-n'-n''}{n' n''}\right| < \frac{n''}{n' n''} = \frac{1}{n'} \sim
\frac{1}{N}\, .
\label{eq40}
\end{equation}
In this situation $n'$ cannot be small compared to $N$ since $n'>n''$ and $n'+n''>N$.
In the case $N<n', n'-n''<N$ the estimate reads
\begin{equation}
\left| \frac{N-n'+n''}{N n''} \right| < \frac{n''}{N n''} = \frac{1}{N} \, .
\label{eq43}
\end{equation}
We see that the cross-correlations (\ref{eq38}) vanish no more slowly
than $1/N$ for $N \to \infty$.

\section{The Form Factor of the Kicked Top}
The Floquet operator of  the kicked top with a chaotic classical limit has proven
faithful  to RMT predictions in many aspects \cite{QSC2}.
Among these, the level-spacing distribution and the integrated two-point correlator
are noteworthy. We here propose to reveal faithfulness of the top to RMT
for the form factor and its fluctuations.

The Floquet operator is built of angular momentum operators
$J_x, J_y, J_z$ and can be understood as a succession of linear rotations and 
torsions. For integer angular momentum values $j$ the operator to be considered
is of the form 
\begin{equation}
F = {\rm e}^{-{\rm i} \frac{\tau_z}{2j+1}J_z^2 - {\rm i} \beta_z J_z}
{\rm e}^{-{\rm i} \beta_y J_y}
{\rm e}^{-{\rm i} \frac{\tau_x}{2j+1} J_x^2 -{\rm i} \beta_x J_x}
\label{eq45}
\end{equation}
and depending on the rotation angles $\beta_i$ and torsion constants $\tau_i$ 
belongs to the orthogonal ($\tau_z = 10, \beta_z = 1, \beta_y = 1,
\tau_x = \beta_x = 0$) or unitary  ($\tau_z = 10, \beta_z = 1, \beta_y = 1,
\tau_x = 4, \beta_x = 1.1$) universality class.
For semi-integer $j$ the Floquet operator
\begin{eqnarray}
F =&& {\rm e}^{-{\rm i} (\frac{\tau_1}{2j+1} J_z^2 + \frac{\tau_2}{2j+1} (J_x J_z + J_z J_x) +
\frac{\tau_3}{2j+1} (J_x J_y + J_y J_x)) } {\rm e}^{-{\rm i}
\frac{\tau_4}{2j+1}
J_z^2} \nonumber\\
\tau_1 =&& 1, \tau_2 = 4, \tau_3 = 2.1, \tau_4 = 10
\label{eq46}
\end{eqnarray}
lacks all geometric symmetries and thus belongs to the symplectic 
universality class\cite{QSC2}.
For numerical work $F$ is represented as a matrix of dimension $N = 2j+1$ in the
basis of eigenvectors of $J_z$, $J_z |j m \rangle = m |j m \rangle $, with fixed
$j$ and $-j \le m \le +j$. 

Figure \ref{fig2} shows the form factors for the three
cases chosen and reveals wild fluctuations about the ensemble means which are displayed 
as the solid lines. The dashed lines together with the abscissa border stripes of 
one and two asymptotic linear variances around the means. 
Gaussian distributions for the $t_n$ would predict the fractions 
$(1-1/{\rm e})/\sqrt({\rm e}) \approx 0.383$ and $(1-1/{\rm e}^2) \approx 0.865$ of all traces to lie
within the one-variance and the two-variances stripes, respectively. 
On counting we find the fractions to be $0.390$ and $0.875$ for the orthogonal top,
$0.397$ and $0.870$ for the unitary one, and finally $0.386$ and $0.858$ for the
symplectic top. The Gaussian expectation is borne out well by these numbers and
further substantiated by the histograms in Fig. \ref{fig3} for the
normalized moduli
$\tau_n = |t_n|/\sqrt{\ovl{|t_n|^2}}$.
Gaussian statistics would yield the distribution $2 \tau
{\rm e}^{-\tau^2}$ which is
displayed as the smooth line and  well approximated by the histograms.

Finally, we illustrate ergodicity of the form factor of the top (unitary variant)
in Fig. \ref{fig4}. The left part shows the form factor fluctuations after a local
time average over the window $\Delta n = 20$. Comparison with Fig. \ref{fig2} reveals the
smoothing effect of the local time average. We have also evaluated the 
variance of the ``band'' of points $|t_n|^2/\ovl{|t_n|^2}$ as a function of the time window
$\Delta n$. The result is represented by the dots in the right part of Fig. \ref{fig4},
together with the $1/\Delta n$ bound predicted by RMT. Again the faithfulness of the
top to RMT is impressive.

\begin{figure}
\caption{The approach of the symplectic ensemble's 
 $\mbox{Var}(|t_n|^2/N) - (\ovl{|t_n|^2}/N)^2$ 
to the asymptotic $- \delta_{|\nu|,1} \pi^2/16$ with growing $N$ 
at half of the Heisenberg time.}
\label{fig1}
\end{figure}

\begin{figure}
\caption{Form factors of the orthogonal, unitary and symplectic  Floquet operators with
parameters as given in the text. Also shown are the respective ensemble means and stripes
of one and two linear variances width around the mean.}
\label{fig2}
\end{figure}

\begin{figure}
\caption{The distributions of the Floqet operators' normalized traces
$\tau_n = |t_n|/\sqrt{\ovl{|t_n|^2}}$ displayed by the histograms show
good agreement with the distribution $2 \tau {\rm e}^{-\tau^2}$ (smooth
line) expected for Gaussian statistics.}
\label{fig3}
\end{figure}

\begin{figure}
\caption{Left: Form factor of the unitary top smoothened by a local time average 
over the window $\Delta n = 20$. Right: Width of the fluctuating band for 
$|t_n|^2/\ovl{|t_n|^2}$ as a function of the time window $\Delta n$ (dots) together with 
the $1/\Delta n$ bound (line) predicted by RMT.}
\label{fig4}
\end{figure}


\begin{thebibliography}{10}
\bibitem{QSC2}
Haake F 1991 {\em Quantum Signatures of Chaos} (Berlin: Springer)
\bibitem{AAA}
Andreev A V and Altshuler B L 1995 {\em Phys. Rev. Lett.} {\bf 75} 902  \newline
Agam O, Altshuler B L and Andreev A V 1995 {\em Phys. Rev. Lett.} {\bf 75} 4389 \newline
Andreev A V, Agam O, Simons B D and Altshuler B L 1996 {\em Phys. Rev. Lett.} {\bf 
76} 1 \newline
Andreev A V, Simons B D, Agam O and Altshuler B L 1996 {\em Nuclear Physics B} {\bf 
482} 536 
\bibitem{FK}
Fishman S and Keating J P 1998 {\em J. Phys. A: Math. Gen.} {\bf 31} L313
\bibitem{BK}
Bogomolny E B and Keating J P 1996 {\em Phys. Rev. Lett.} {\bf 77} 1472
\bibitem{SMI1}
Smilansky U  1997 {\em Physica D} {\bf 109} 153
\bibitem{SMI2}
Kettemann S, Klakow D and Smilansky U 1997 {\em J. Phys. A: Math. Gen.} {\bf 30} 3643
\bibitem{PRA}
Prange R E 1997 {\em Phys. Rev. Lett.} {\bf 78} 2280
\bibitem{KUN}
Kunz H and Shapiro B 1998 {\em Phys. Rev. E} {\bf 58} 400 \newline
Kunz H 1999 {\em J. Phys. A: Math. Gen.} {\bf 32} 2171
\bibitem{HOdA}
Hannay J H and Ozorio de Almeida A M 1984 {\em J. Math. Phys. A} {\bf 17} 3429
\bibitem{PAN}
Pandey A 1979 {\em Ann. Phys.} {\bf 119} 170
\bibitem{HKS}
Haake F, Ku\'s M, Sommers H-J, Schomerus H and \.Zyczkowski K 1996 
{\em J.Phys. A: Math. Gen.} {\bf 29} 3641
\bibitem{SHW}
Sommers H-J, Haake F and Weber J 1998 {\em J. Phys. A: Math. Gen.} {\bf 31} 4395 
\bibitem{DYS}
Dyson F J 1962 {\em J. Math. Phys.} {\bf 3} 140 
\bibitem{MET}
Mehta M L 1991 {\em Random Matrices} (New York: Academic) 


\end{thebibliography}
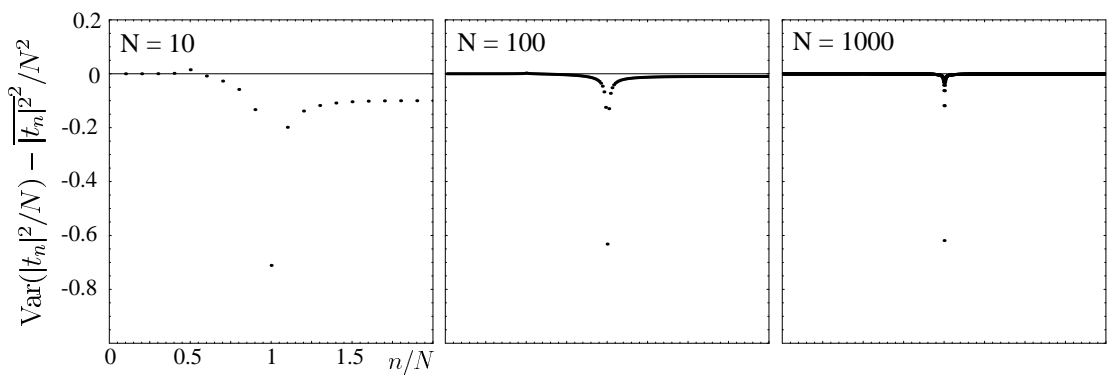
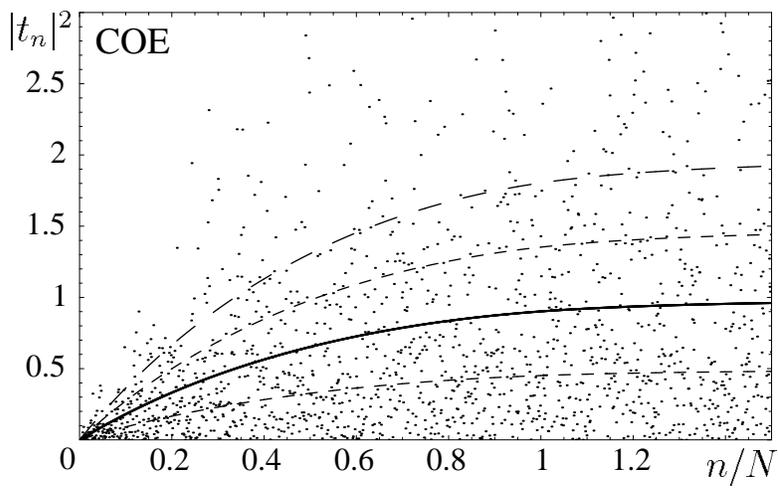
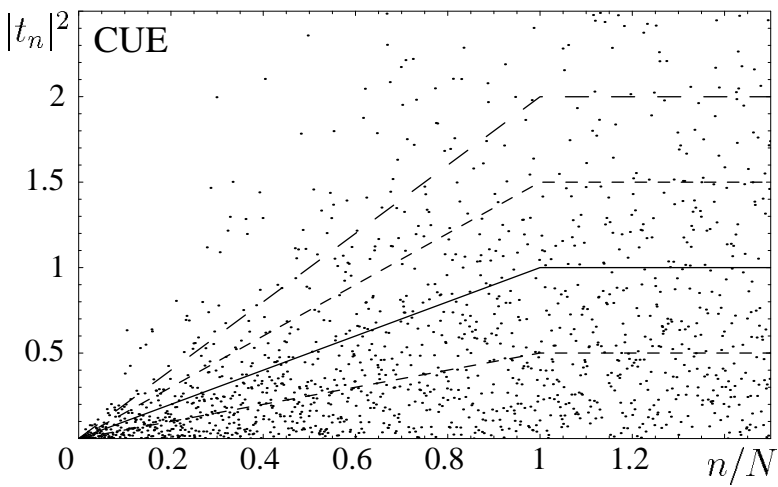
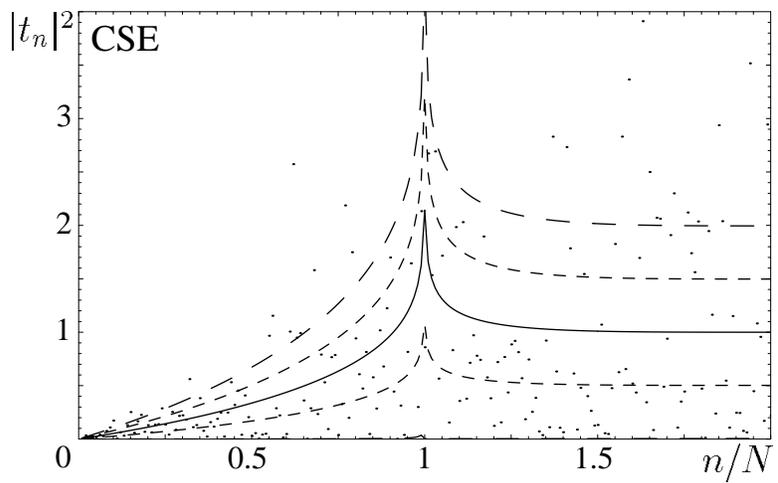
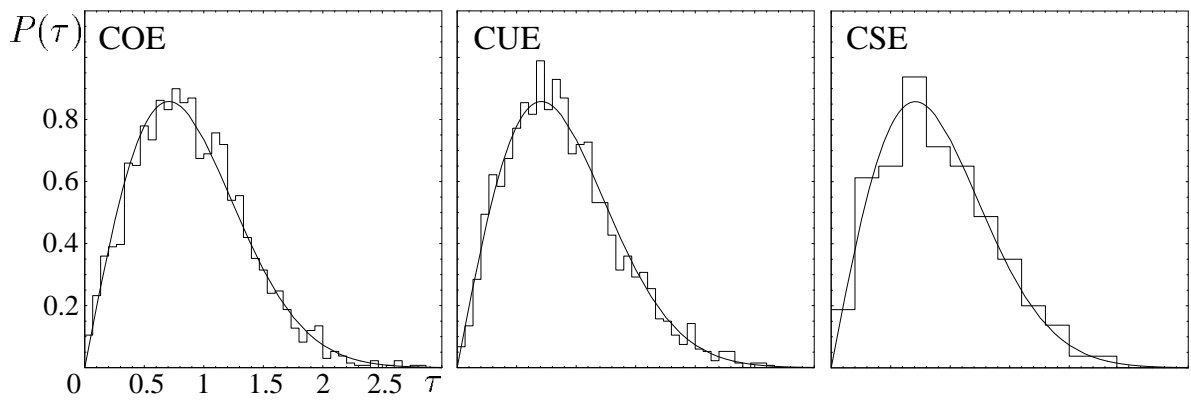
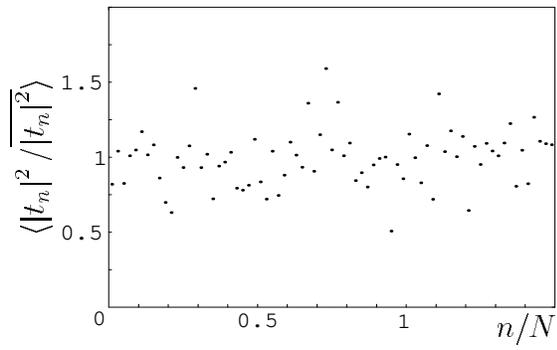
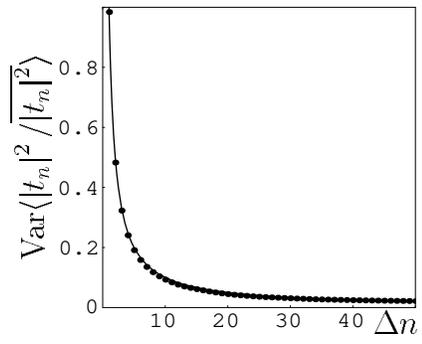
\end{document}